# A Study on the Possible Effects of the Implementation of the Nordic Model in India on Crime Rates and Sexually Transmitted Diseases

[1]Sabarinath Vinod Nair, [2]Shreya Sharma, [3]Swarnava Ghosh

[1]Department of Statistics,
[1]CHRIST (Deemed to be University), Bengaluru, India

[1]sabarinaathnair@gmail.com, [2]shreyasharma7612@gmail.com, [3]swarnava98@gmail.com

*Abstract*—**Prostitution is one of the root causes of sex trafficking and the transmission of sexual diseases. The rules and regulations followed by the Indian government to regulate the same, fall under the umbrella of the abolitionism model. Neo-abolitionism (also known as the Nordic model) is a new legislative model that has been introduced by the Nordic countries to regulate prostitution. The purpose of this research paper is to examine the possible effects of the application of the Nordic model on the crime rates and the spread of sexually transmitted diseases in India. Further, we also aim to study the effects of the implementation of Neo-abolitionism in Sweden.**

## I. INTRODUCTION

Prostitution is one of the oldest professions in the world, not just in India, dating back to the start of the organized civilization [1]. In Europe, during the middle Ages, church leaders attempted to rehabilitate penitent prostitutes and fund their dowries. Nevertheless, prostitution flourished; it was not merely tolerated but also protected, licensed, and regulated by law, and it constituted a considerable source of public revenue. Public brothels were established in large cities throughout Europe. At Toulouse, in France, the profits were shared between the city and the university. In England, bordellos were originally licensed by the bishops of Winchester and subsequently by Parliament. Geishas in Japan were not only a respected profession but they were sought after as performers, artistes, concubines and even wives for the samurai.

Prostitution mainly refers to sex for money, but over the centuries its definition, approach and purpose have changed. The erstwhile glamorous profession in India has now taken an unusually derogatory form. Not only are they looked down upon, but they are also treated as outcasts from society. This has led to the sex industry catering mostly to the uneducated, the impoverished and the goons. As a result, organized prostitution has become one of the root causes of the spread of sex trafficking, sexually transmitted diseases and other vital vices that have crept into society. Quite unlike the olden times, people who enter this trade now are either forced into it via trafficking or their poor financial conditions have not left them with any other stable alternative.

In the current scenario, prostitution has been defined in different ways depending on the extent of its prevalence. The Prevention of Immoral Traffic Act (1987) by the Government of India defines the term "prostitution" as "sexual exploitation or abuse of persons for commercial purposes." Nevertheless, the most widely accepted definition is the one given in the encyclopedia of social science which defines prostitution as "the practice in which a female offers her body for promiscuous sexual intercourse for hire, etc." Evidently, male prostitutes, though fairly prevalent, are not as stigmatized as the females, in the present society.

As a part of the initial five-year plan in independent India, a survey on sex workers was conducted [2]. Based on the survey, 79.4% of women entered prostitution due to poverty, while rest was either trafficked, cheated, sold or were devadasis. (Devadasis are girls who had been offered by their parents to the service of God in temples. They are supposedly married off to the deity of the temple.) Though originally supposed to be dancing during prayer songs, due to the laxity of morals among the priests, misuse of power by village heads and others in power in the villages, these girls became their private prostitutes, resulting in a clandestine umbrella of prostitution garbed in the name of religion and offering to God. [1]

In 2007, the Ministry of Women and Child Development, Government of India, [3] reported the presence of over 3 million female sex workers in India, with 35.47 percent of them entering the trade before the age of 18 years. As per an estimate by UNAIDS [4], there were 657,829 prostitutes in India in the year 2016 [5]. "Every hour, four women and girls in India enter prostitution, three of them against their will".

The tourism industry is developing at a rapid pace and has played a major role in providing a livelihood to many people including hoteliers, tour operators, and travel guides. It even directly contributes to the economy of the country. However, on the negative side, it has made way for a significant increase in child sex tourism. One of the main reasons behind this is the mentality of the tourist of being free from the social and legal constraints of their own country. The travellers who come to India for fulfilling their sexual desires include a category of people termed as pedophiles. Pedophiles refer to the people who prefer to copulate with minors for the fulfillment of their carnal desires. Some of the major factors affecting child sex tourism are [6]:

i) The feeling among foreign tourists that children of third world countries can be exploited and that the chances of

detection are slender.

 ii) A belief that children are less likely to have contracted sexually transmitted diseases and hence fornicating with them is safe.

 iii) The mistaken notion that fornicating with virgin girls cures HIV.

 iv) Lax enforcement of the law in third world countries and even lack of strict punishments as compared to western countries on child trafficking.

 v) Availability of children, especially in Muslim dominated areas, for sham marriages, sale and even in the garb of employment to be carried away to gulf countries by the rich for sexual exploitation and abuse.

 vi) The governments of many developing countries, with a view to encouraging tourism, turn a blind eye to this problem.

Whether prostitution is empowering, exploitative or a consequence of morality is an entirely subjective argument, but the political regulation of prostitution differs internationally, owing to every country's respective economic and social outlook. Some of the main legislative responses to prostitution in specific social constructs can be categorised as (1) Criminalisation, (2) Legalisation / Decriminalisation, and (3) the Nordic Model [7].

The Nordic model of prostitution policy originates from the social democratic theory, Marxism and radical feminism. It interweaves the best of both worlds from the criminalisation and legalisation/decriminalisation frameworks to provide an approach more consistent with social democratic tendencies. It aims to use the government and social policies to support the equality of all human beings. The model neither considers prostitution as labour nor a result of the immorality of women. It instead conceptualises prostitution as an act of gender violence [7]. It decriminalises people who are prostituted and criminalises the buyers of sexual services and third-party profiteers (pimps and brothel owners), thereby providing safe passage to the prostitutes to try and exit the trade that they were forced into. However, this model might come across as one sympathising with women (from quite a gender-specific perspective) as victims of the trade, but it is not really irrelevant in the context of India, as is in Sweden. Main highlights of the Nordic model for prostitution:

 i) Decriminalises prostitutes
 ii) Helps them exit prostitution
 iii) Makes sex trafficking an offence
 iv) Countries which have already implemented the Nordic model are Sweden, Norway, Iceland, Northern Ireland, Canada, France, Ireland and Israel.

Model's main approach:
i) The full decriminalisation of those who are prostituted.

 Evidence suggests that the majority of women and children enter prostitution as a result of childhood abuse, poverty and misfortune, grooming, coercion, and/or betrayal, rather than as a free choice between a numbers of viable options.
The evidence clearly establishes that prostitution is inherently violent and damages those in it and that getting out of it is much harder than getting into it and a criminal record makes getting out even harder. The management of prostitution by pimps and enforcers is intrinsically violent and depends on instilling fear and pain as disincentives to those that contemplate leaving. The fact that in most countries, the law enforcement agencies are corrupt and complicit leaves the victims very little choice and very few even dares to try getting out. Therefore, all the laws that target those who are prostituted and the clearing of their criminal records of any previous convictions for offences related to their own prostitution is a necessary first condition to encourage free will as a condition of continuation.

ii) High-quality services for those in prostitution.

 Funding should be provided for high-quality social and medical services for those already in prostitution. These must be non-judgemental and cover harm reduction as well as exit support, including housing, legal advice, de-addiction services, long-term emotional and psychological support, education, training and childcare.

iii) Buying sex to be made a criminal offence.

 We call for the purchase and attempted the purchase of human beings for sex to be made a criminal offence, regardless of where in the world it takes place. We do not believe foreign men should be free to cause damage in other countries. As explained earlier, the aim is to change behaviour rather than to criminalise people. We recommend a maximum sentence of one year in prison.

iv) The procuring, pimping and sex trafficking legislation to be strengthened.

 We believe that the current pimping and sex trafficking legislation is not fit for purpose and we call for it to be replaced with stronger legislation that recognises procuring, pimping and sex trafficking as the human rights abuses that they are and for penalties that they reflect. The policing of these crimes must be fully resourced and prioritised.

v) All the factors that drive people into prostitution to be addressed

 Prostitution cannot be accepted as the answer for the poor and disadvantaged, for single mothers, for women and children or indeed for anyone. Therefore, there is a call for a fairer and more equal society with a guaranteed minimum income for all, the elimination of the pay gap between women and men, better resources and support for parents and "looked after" children, an end to student fees and zero-hour contracts, and tackling of all the other factors that trap people in poverty.

vi) A holistic approach

To be effective, the Nordic Model must be accompanied by a widespread public information campaign (like the one that accompanied the change in the smoking laws). Education programmes in schools that explain honestly the damage that prostitution causes. Experience in other countries has shown that for the Nordic Model to be effective, it needs to be accompanied by in-depth training for the police, judiciary, Crown Prosecution Service (CPS), and frontline workers in education, social services, local government, the NHS, etc. For the Nordic Model approach to be effective it needs to be prioritised and implemented consistently across the country, otherwise, pimps and punters will simply move to areas where it is not enforced. Similarly, services for those who are prostituted must be coordinated nationally and not be left to the localism agenda.

The abolitionist model currently in practice in India criminalises both the buyers and the sellers of sexual services. This evidently does not do justice to the women who are mostly forced, trafficked, sold or tricked into entering the trade. However, the Nordic model, if implemented in the Indian context can try and solve the problem faced by numerous such women who are being exploited in the industry and holds the power to regulate and eventually eradicate the practice of buying sexual services. As a result, several criminal activities strongly dependent on the existence and practice of prostitution can also be minimized. In this paper, we primarily aim to examine the effect of the application of the Nordic model in use in Sweden, on crime rates and the spread of sexually transmitted diseases in India.

## II. OBJECTIVES

1. Understanding the rate of rapes and STI's and STD's in Sweden and the factors influencing it and how these factors influence rapes in India.

2. To obtain a model that defines the after-effects of implementation of the Nordic model in Sweden on sex trafficking, rape and spread of HIV diseases.

3. Understanding the state of human trafficking in both countries.

4. To find the difference that would occur in number of rapes, sex trafficking and spread of HIV diseases upon the implementation of the Nordic model on prostitution in India.

## III. REVIEW OF LITERATURE

To gain qualitative clarity of the topic under review, numerous theoretical and analytical research articles pivoting about prostitution were considered. The study is based on a fundamental understanding of the legal frameworks of the countries, the similarity and the differences in the social standpoint and the economic status of the population.

Prostitution is not technically illegal in India. No law states that prostitution itself is illegal, but the core activities associated with it are. These activities include but are not limited to owning and running brothels (which is any place with more than two sex workers), kerb-crawling, soliciting sex in a public area, seeking sexual favours for money in public, pandering and illicit activities in hotels. This places an almost complete restriction on and around this field of work.

The fundamental law regarding a prostitute's status in the society, The Immoral Traffic (Suppression) Act, was passed in 1956. It is commonly abbreviated and referred to as SITA. This act states that sex workers are allowed to carry out their trade-in private but they cannot ply their business in the open. The Indian laws, however, do not regard buying or selling sex as prostitution [7]. As per the laws, the buyers can be arrested if they are found guilty of indulging in any sexual activity in a public place. Even though buying and selling sex is permissible on individual capacity, a woman cannot do it within a span of 200 yards of a public place. Sex workers do not fall under the umbrella of general labour laws. However, they possess all the rights that would be enjoyed by a citizen and are entitled to be rescued and rehabilitated if they want to do so. It is probable only if society would accept the possibility of such activity [8].

Nevertheless, SITA is not used as such. Mostly, different sections of the IPC are employed to bring charges of long drawn criminal acts such as public indecency to get back at the sex workers. They can also be accused of being 'public nuisance' under IPC. The most basic problem is that there is no clear definition of what these crimes constitute and sex workers are basically left to the mercy of the officials who bring the charges against them.

SITA has recently been changed to become PITA or The Immoral Traffic (Prevention) Act. There have been several attempts to change this law so that a bigger slice of the blame can be placed on the clients. However, the Union Health Ministry has opposed such developments [9]. These days, insurance companies are coming forward and insuring sex workers. This is a very welcomed move. Sections 372 and 373 restrict the sale and purchase of minors for the purpose of prostitution and illicit intercourse. This does not do much good as the culprits knew it was morally wrong even if it wasn't explicitly mentioned to be illegal, but continue to exploit womanhood. According to Apne Aap, an NGO fighting against human trafficking says that around 30 percent of the women trafficked are minors.

The sad thing is even when these little, innocent girls are rescued, they are again sold to the same middlemen by their own parents. There are no strict laws as to effectively stop the culprits once they have been convicted of the crime [10]. This law does not protect the rights of the sex worker but criminalises the activities related to prostitution. Prostitution is legal in India but the activities are not. The demand for legalisation of prostitution in India is really high but some of them consider this against the norms of the Indian society. In a country with over 2 million sex workers, legalizing their occupation would mean that this industry would be a major source of tax revenue. Moreover, it would help in improving the health and living standards of the thousands of women who spend their lives deprived of all civil opportunities that are available outside the "red-light" areas.

If prostitution is legally recognised as a valid occupation, and sex workers as normal wage workers, a lot of shortcomings of the business can be tackled. Being able to legally run their services without the fear of being victimised would mean that the sex workers would not need to rely on middlemen or brothel owners who more than often personify cruelty [8]. They would be the owners of their own body and rights, which is what everyone else but them enjoys. The right to have control over one's body is the most basic of all human rights, and shouldn't even need mentioning. While prostitution cannot and need not be maintained, it needs to be regulated.

## IV. METHODOLOGY

This paper performs a deductive comparative analytical research on the possible effects of the implementation of the Nordic model currently in use in Sweden, in the case of India, with respect to prostitution. The major factors influenced by prostitution that have been considered are human trafficking, use and spread of narcotics, the spread of sexually transmitted diseases and infections, asylum migrations, tourism data, rape cases and rate of unemployment in both the countries. All of the findings in this paper are based on secondary data collected directly or indirectly from government websites of the respective countries. Time series data have been considered to analyse the change in the numbers before and after the implementation of the legislative model. To decide the factors required for study, several theoretical research papers on prostitution were consulted to gain a qualitative idea of the general factors influencing prostitution [1] [6] [7] [12]. For the effectiveness of the quantitative analysis and ease of visualisation, the raw data was collected for a total number of cases registered in a year, converted into cases/parts per million of the population and then considered for analysis.

Mainly, correlation and regression analyses have been used to compare and relate the factors under study. The analysis involves developing simple and multiple linear regression models with a combination of the factors under consideration as regressors and response variables and fitting the same model for India, by feeding relevant data. To test which of the factors (regressors) significantly affect the dependent variable in the model, both forward substitution and backward elimination methods of variable selection have been used. The AIC (Akaike Information Criteria) and the ANOVA models have been considered for the variables to test the quality and accuracy of the model and the variables have been accordingly chosen for the study. The t-test has been used for testing the hypotheses and significance. The time-series data collected from secondary sources have been utilised to provide better visualisation of graphs.

Most of the analysis and relevant visualisations have been done in RStudio, using the R programming language. However, some of the direct visualisations of the data have been performed in Microsoft Excel. The limitations of the methodology used revolves around the accuracy of the data as it is mostly collected from secondary sources. Only the cases registered by the government have been considered. However, the actual number of cases, many of which go unreported, may be different. Also, the data of criminal activities are often tampered and manipulated. Hence, it is difficult to ascertain the accuracy of the analysis. However, it is practically not viable to get the primary data on these factors as the safety of the survey personnel remains at stake in the circumstances of data collection.

## V. ANALYSIS

### A. Rapes vs. Asylum Migration

As per multiple reports, the vital reason behind the increase in the number of rapes is an increase in the number of immigrants. It is found that both the factors are directly proportional and hence the trend. In 2018, Swedish Television investigative journalism show Up drag Granskning analysed a total of 843 district court cases from the five preceding years and found that 58% of all convicted of rape had a foreign background. On considering the relevant factors, ranging from 1992-2017 it was observed that there exists a negative correlation of -0.169 between the number of rapes and asylum migrants per million (*Figure 1*). The high inflow of migrants was due to a result of communism breakdown in Eastern Europe in the year 1989. Moreover, the civil war in the former Yugoslavia also resulted in a stream of refugees to neighboring countries. Sweden received more than 80,000 refugees in 1992, the majority (almost 70,000 persons) coming from the former Yugoslavia.[14] There was an increase in the number of crimes after the influx of migrants and refugees were found guilty in most of the cases. Along with this, there is a general fear in Sweden that crime will enter from the countries around the Baltic Sea, which might be another reason for the belief in the fact that high level of immigrants is a major factor responsible for the increased number of rapes. Similarly, in India during the same time period, -0.5 is the value obtained. The value, in turn, indicates that these two factors have the least dependence on each other (*Figure 2*).

*Table 1. Rape Rate and Asylum Migration (per million)*

| Year | Rapes | Asylum Migration |
|---|---|---|
| 1992-1997 | 204 | 238 |
| 1997-2002 | 222 | 94 |
| 2002-2007 | 341 | 121 |
| 2007-2012 | 614 | 141 |
| 2012-2017 | 655 | 197 |

*Figure 1. Sweden (per million)*

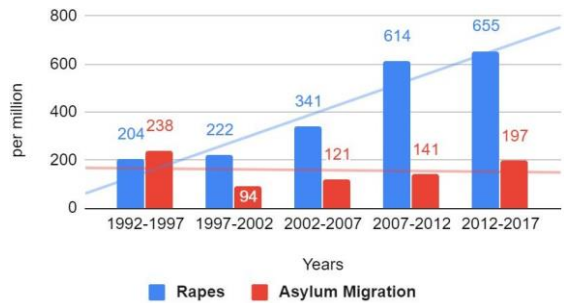

*Table 2. Rape Rate and Asylum Migration (per million)*

| Years | Rapes | Asylum Migration |
|---|---|---|
| 1992-1997 | 22 | 262 |
| 1997-2002 | 21 | 179 |
| 2002-2007 | 24 | 140 |
| 2007-2012 | 28 | 148 |
| 2012-2017 | 29 | 150 |

*Figure 2. India (per million)*

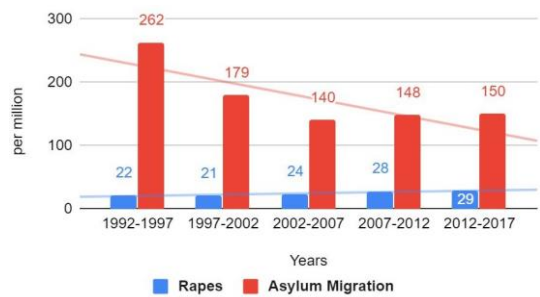

India and Sweden:

As per the data accumulated from a trusted third party website and our data as well, Sweden accounts for more number of cases than India and has a higher global ranking in rape rates. One of the main reasons that can be considered for this is the case is the stigma associated with sex and rape and difference in the legal framework of the two countries.

*Table 3. Global Rape Rate Ranking*

| Crime | India | Sweden |
|---|---|---|
| Rape rates | 1.8<br>Ranked 46th. | 63.5<br>Ranked 3rd. 35 times more than India |

*Figure 3. India vs. Sweden (Rape rates)*

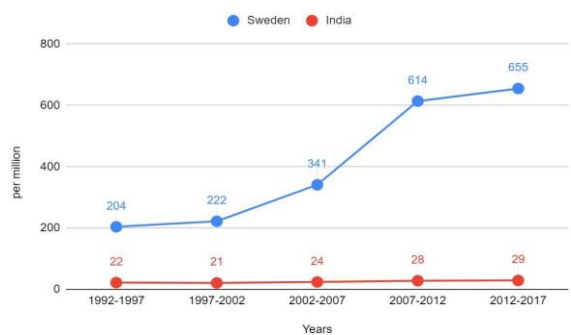

Though *Table 3* and *Figure 3* clearly suggest that Sweden ranks higher than India in rape rates, yet the fact that Sweden has been way ahead of India in terms of taking actions for gender equality and sex education cannot be ignored. In 1965, Sweden was one of the first countries in the world to criminalise marital rapes [16]. Homosexual acts and gender neutrality was first introduced in 1984, [17] and sex with someone by improperly exploiting them while they are unconscious (e.g. due to intoxication or sleep) was included in the definition of rape in 2005. [18]

Along with this, in the year 1954 Sweden became the first country to introduce sex education in the year. The findings of the 2000 International Crime Victims Survey (ICVS) [19] indicate that the respondents' satisfaction with the police is above average in Sweden, with almost no experience of corruption. [20] Sweden has also been ranked number one in sexual equality. In a world where crimes are not tabooed and corruption is seldom found in legal bodies, people have faith in the judiciary and feel free to report cases. Such a situation is obviously utopian from the perspective of an Indian.

In India, due to the stigma attached to the crime and comparatively higher corruption rates, many cases go unreported. Indian parliamentarians have stated that the rape problem in India is being underestimated because a large number of cases are not reported, even though more victims are increasingly coming out and reporting rape and sexual assaults [18] [21].

Few states in India have tried to estimate or survey unreported cases of sexual assault. The estimates for unreported rapes in India vary widely. The National Crime Records Bureau report of 2006 mentions that about 71% rape crimes go unreported [22] Marital rape is not a criminal act in India [23] though sexual intercourse with a wife aged between 15 and 18 years is considered as rape [24].

### B. Effects of Implementation of Neo-abolitionism on Rape Cases

For this particular test we have taken data in cases per million recorded from the year 1978 - 2018, to compare the check whether there was any change in the values before the new model was defined in Sweden and after it was defined.

The model was implemented in the year 1999, so the values corresponding to the year 1978 - 1998 have the values before the model was implemented and values corresponding to the year 1999 - 2018 have the values after the model was implemented.

Formation Of Hypothesis:

$H_0$: There is no significant difference in the rape cases after the implementation of neo-abolitionism.

$H_1$: There is a significant difference in the number of rape cases after the implementation of neo-abolitionism.

To understand the change in rapes committed before and after the implementation of Neo-abolitionism, the paired samples t-test has been used, the results of which are tabulated as follows:

*Table 4. Paired Sample Test Results*

| | Paired Differences | | | | | t | df | Sig. (2-tailed) |
|---|---|---|---|---|---|---|---|---|
| | Mean | Std. Dev. | Std. Error Mean | 95% Confidence Interval of the Difference | | | | |
| | | | | Lower | Upper | | | |
| Pair 1 Before - After | -34.000 | 175.546 | 45.326 | -131.214 | 63.214 | -.750 | 14 | 466 |

Here from the *Table 4*, the p-value obtained is less than 0.05, which in turn indicates that there is a difference in the number of rapes before and after the implementation of Neo-abolitionism in Sweden. Also, the difference between the average of rape rates before and after the implementation of the nordic model is 34, which indicates an increase in the number of rapes after the implementation of neo - abolitionism. The trend of change of rape cases (Figure 3) in the case of Sweden, clearly indicates an increase in the number of rape cases. Increase after the year 1999 can be related to the implementation of the Nordic model. According to the model, whosoever buys sex is criminalised, which in turn can lead to an increase in the number of rape cases. As people who did not have a partner resorted to this method.

This increase can also be related to increased access to pornography in the digital world. Though Sweden has a proper provision for sex education, people still get influenced by what they watch. As per many psychological researchers, watching porn extensively distorts their perception about sexual intercourse. Rather than looking at the gender of interest as human, they start objectifying them. Increase in cases in the year 2006 can also be related to the increase in the number of refugees or asylum migrants and tourists in Sweden. Yet, it is undecided whether these people who come from outside have a criminal mindset or rather they do not find any other alternatives to fulfill their needs.

### C. Mean Age of Marriage and Total Sexually Transmitted Diseases'

Sweden:

At a time of life when adolescent males and females are just beginning their sexual decision-making, they are at greatest risk for contracting sexually transmitted diseases (STD), including HIV [25]. Sexually transmitted disease being the most encountered diseases across the globe is an important factor under consideration in this research paper too. It is considered as one of the independent variables in the further part of the analysis with which a model is fitted with respect to both the countries and we try to check the compatibility of the model. Here in the correlation analysis, we have considered the mean age of women and men when they give birth to their first kid over the years and the relationship between the diseases that are prone to occur. So, here we obtain a strong positive correlation in case of both men and women and it is quite evident that the diseases can affect any of them irrespective of their gender and necessary precautions should be taken to prevent the occurrences of these diseases.

*Table 5. Age of getting married for men and women, and total STD's and STI's*

| 0.8273104777 | Correlation between the mean age at which men get married and total STD and STI |
|---|---|
| 0.7787088547 | Correlation between the mean age at which women get married and total STD and STI |

India:

India also shows substantial diversity in age at marriage, particularly for very early marriage. National Family Health Survey III documented that even among recent cohorts of women aged 18 -29, 52% of the women were married by age 18 in Uttar Pradesh; corresponding proportions were 25% in Tamil Nadu and 17% in Kerala [26] [27]. For this research, we have taken the average age of marriage of men and women for the whole country. Similarly for India also, we calculate the correlation between the total number of cases for STDs and the average age of marriage for men and women in India.

*Table 6. Age of getting married for men and women, and total STD's*

| 0.5534796351 | Correlation between total STD and the average age of marriage of men |
|---|---|
| 0.6621788095 | Correlation between total STD and the average age of marriage of women |

There is a moderate correlation between rape cases and the average age of marriage. This indicates that the age of marriage does affect the spread of STDs/STIs in India to a considerable degree. The reason is the same as stated for Sweden, i.e. the probability of having number of partners increases and in turn, it leads to an increase in the probability of getting infected by such a disease. Even after multiple awareness programs by the government and NGOs, people refrain from discussing the importance of contraceptives and sanitary napkins. The stigma attached to this is one of the major causes behind the spread of disease.

*D. Fitting of a Model*

The setting up of the model involves developing simple and multiple linear regression models with a combination of the factors under consideration as regressors and response variables and fitting the same model for India, by feeding relevant data. In the first place, we consider the independent variables as Asylum migration, Drug crimes and Tourism. All these factors are directly related to the level of the widespread of STI/STDs in a country. Tourists when they enter a third world country might want to fulfil their sexual desires which might not be possible for them in their own country. Immigrants have a high probability of carrying the traits STD/STIs from their native country. People who consume drugs usually end up injecting using the same sterile needles or syringes. These regressors are then pair-wisely compared with the dependent variable that is sexually transmitted diseases. All the relevant data are considered in a pairwise manner with respect to the corresponding total of sexually transmitted diseases. All the relevant data over the past so many years are gathered together and the backward variable selection criteria are carried on the independent variables with respect to the dependent variable. Based on the Akaike Information Criteria, lower the value better is the regressor in explaining the dependent variable. Or in other words, we can say that the particular regressor affects the dependent variable significantly.

*Table 7. Linear Regression Coefficients for the Swedish Model*

| Coefficients | Estimate | Std. Error | t value | Pr (>|t|) |
|---|---|---|---|---|
| (Intercept) | 330.2466 | 71.57403 | 4.614 | 0.000168 |
| Asylum Migration | 0.30257 | 0.15802 | 1.915 | 0.069945 |
| Tourism | 7.75949 | 1.56281 | 4.965 | 7.45e-05 |
| Drugs | 0.09858 | 0.14975 | 0.658 | 0.517843 |

This procedure is firstly carried out on the parameters of Sweden and we obtain the following result. From the statistical summary of the model, we can conclude that the adjusted Coefficient of Determination (adjusted R-squared) value is 0.6119 and says that the model is fairly good.

Further, we determine the regressors that are significantly affecting the dependent variable, that is, sexually transmitted diseases. Asylum migration and tourism affect the dependent variable, as they show the maximum effect on the dependent variable. Similarly, we carry out the same procedure for the Indian figures with the same as that of Sweden.

*Table 8. Linear Regression Coefficients for Indian Model*

|  | **Estimate** | **Std. Error** | **t value** | **Pr(>|t|)** |
| --- | --- | --- | --- | --- |
| Intercept | -104.851 | 157.26534 | -0.667 | 0.51257 |
| Asylum Migration | 2.01784 | 0.55579 | 3.631 | 0.00167 |
| Tourism | -0.02562 | 0.02309 | -1.109 | 0.28041 |
| Drugs | 1.59693 | 3.34463 | 0.477 | 0.63821 |

In India, the influence of asylum migrants and drugs on the spread of STD's is comparatively higher than Sweden. Tourism is comparatively lesser. The adjusted R squared value obtained for the above model is 0.6034 which makes this as a moderately good model for India for the given scenario.

## VI. CONCLUSION

This paper has attempted to understand the implications of Implementation of Neo-Abolitionism with respect to their impact on Sexually Transmitted Diseases and rapes. In order to completely evaluate the same, and their relationship with each other, various factors present behind them in Sweden are analysed and then the same factors are used for India as well. The first factor that was evaluated to understand the after-effects of Nordic Model was the rate of rapes in the respective countries. As per many theories and reports, the crime rates increased in Sweden after the inflow of migrants. We analysed the influence of asylum migrants on rapes for both the countries. We found that there was no dependency of the number of migrants and rapes that were reported in Sweden but a moderate dependence was found in case of India. The rate of rapes effectively doubled after the year 1999 in Sweden and this can be related to the implementation of Neo-Abolitionism. According to the model, whosoever buys sex is criminalised, which in turn can lead to an increase in the number of rape cases.

In case of Sexually Transmitted Diseases and Infections, it was found that the asylum migrations and tourists were a major cause of the spread of the various diseases and infections in Sweden as compared to any other factors in Sweden. The same model when tried on the Indian the data showed similar results. This, in turn, shows that India might have the same effect on STDs with respect to the same factors that were taken into consideration for analysing the data for Sweden.

On a whole, we can conclude that there was no improvement in Sweden in case of rape rates and spread of STDs after the implementation of Neo-Abolitionism. Rather the rates increased, which throws light on the unregulated prostitution that currently prevails in the country. Nordic model in case of prostitution is just a part of the legalisation of prostitution which can again leave a segment of the market unregulated and hence gives rise of corruption and black markets.